\documentclass[aps,pra,superscriptaddress,twocolumn,showpacs,preprintnumbers,amsmath,amssymb,natbib,balancelastpage,longbibliography]{revtex4-1}
\usepackage[dvips]{graphicx}
\usepackage{srcltx}
\usepackage{color}
\usepackage{bm,upgreek}
\usepackage{amsmath,mathtools}
\usepackage{amsfonts}
\usepackage{amssymb}
\usepackage{mathrsfs}
\usepackage{boxedminipage}
\usepackage{framed}
\usepackage{bbm}
\usepackage{natbib}\usepackage[ugly]{nicefrac}
\usepackage{mathtools}
\usepackage{enumitem}
\usepackage{multirow}

\def\del#1{\color{grey}\sout{#1}~\color{black}}

\def\bra#1{{\langle #1|}}                     
\def\ket#1{{|#1\rangle}}                      
\def\braket#1#2{{\langle #1|#2\rangle}}       
\def\ketbra#1#2{{|#1\rangle\langle#2|}}       
\def\matel#1#2#3{{\bra{#1}#2\ket{#3}}}        
\def\comm#1#2{{\big[#1,#2\big]}}              
\def\mod2#1{{\big|#1\big|^2}}                 
\def\av#1{{\langle#1\rangle}}                 
\def\adj#1{{#1}^\dagger}                    
\def\set#1{{\left\{#1\right\}}}               
\def\-{\!-\!}                               
\def\+{\!+\!}                               
\def\={\;=\;}                               

\def\A{\op{A}}
\def\Fc{I}
\def\Fq{J}
\def\H{\op{H}}

\def\V{\op{V}}

\def\Q{\op{Q}}

\def\0{\vec{0}}

\def\d{\mathrm{d}}                          
\def\intd#1{\int\d#1\;}                     
\def\intd2#1{\int\d^2#1\;}                  
\def\intd3#1{\int\d^3#1\;}                  
\def\del{\partial}                          
\def\vec#1{\mathbf{#1}}                     
\def\op#1{\hat{#1}}                         
\def\set#1{\{#1\}}                          
\def\-{\!-\!}                               
\def\+{\!+\!}                               

\def\var#1{{\mathrm{var}[#1]}}

\def\beq{\begin{equation}}                  
\def\eeq{\end{equation}}                    

\def\Angstrom{\AA ngstr\"om}

\begin{document}


\title{Quantum and Classical Fisher Information in Four-Dimensional\\ Scanning Transmission Electron Microscopy}

\author{Christian Dwyer}
\email{dwyer@eistools.com}
\affiliation{Electron Imaging and Spectroscopy Tools, PO Box 506, Sans Souci, NSW 2219, Australia,}
\affiliation{Physics, School of Science, RMIT University, Melbourne, Victoria 3001, Australia}

\author{David M. Paganin}
\email{david.paganin@monash.edu}
\affiliation{School of Physics and Astronomy, Monash University, Clayton, Victoria 3800, Australia}


\begin{abstract}

We analyze the quantum limit of sensitivity in four-dimensional scanning transmission electron microscopy (4D-STEM), which has emerged as a favored technique for imaging the structure of a wide variety of materials, including biological and other radiation-sensitive materials. 4D-STEM is an indirect (computational) imaging technique, which uses a scanning beam, and records the scattering distribution in momentum (diffraction) space for each beam position. We find that, in measuring a sample's electrostatic potential, the quantum Fisher information from 4D-STEM can match that from real-space phase-contrast imaging. Near-optimum quantum Fisher information is achieved using a delocalized speckled probe. However, owing to the detection in the diffraction plane, 4D-STEM ultimately enables only about half of the quantum limit, whereas Zernike phase-contrast imaging enables the quantum limit for all spatial frequencies admitted by the optical system. On the other hand, 4D-STEM can yield information on spatial frequencies well beyond those accessible by phase-contrast TEM. Our conclusions extend to analogous imaging modalities using coherent scalar visible light and x-rays.
\end{abstract}

\maketitle 

\section{Introduction}

Quests persist to develop ever-more sensitive imaging techniques to probe the structure of materials down to the atomic level. Ultra-high sensitivity becomes absolutely mandatory when studying samples such as quantum materials and radiation-sensitive materials. For radiation-based imaging techniques, dose efficiency is of primary importance, i.e., for a given ``radiation budget," what precision can be achieved in measuring the material's properties of interest? All quests for sensitivity/precision are ultimately bound by the laws of quantum mechanics. Such limitations are most famously known in the form of the Heisenberg uncertainty relations. However, there now exists a considerably more general formalism known as quantum estimation theory, which can offer significant insights into the limitations of a given technique, and provide reasons why certain techniques enable maximum dose efficiency.

Here, building on recent work \cite{Dwyer2023}, we apply the formalism to analyze the sensitivity of (diffraction-based) imaging in four-dimensional scanning transmission electron microscopy \cite{Ophus2019} (4D-STEM), which has emerged as a favored technique for imaging the atomic structures of a wide variety of materials, including radiation-sensitive materials. In this technique (in fact, class of imaging modes), an electron beam is scanned across the sample, and for each beam position the distribution of scattering is captured by a pixellated detector in momentum (diffraction) space, producing a four-dimensional dataset. The images are reconstructed via a computational algorithm, which can range from simple to quite extensive, using the dataset as input.  The technique permits a range of imaging modes, several of which are capable of deep sub-\Angstrom\ resolution (especially in the case of ptychography) and sensitivity to both light and heavy elements.

Our quantum estimation theory-based analysis reveals that, when optimized, 4D-STEM can attain about half of the available quantum Fisher information, meaning that, for a given level of precision, it requires about twice the minimum electron dose permitted by quantum mechanics. For an arbitrary spatial frequency, we find that near-optimum information transfer is achieved by a delocalized speckled probe. Preclusion of the quantum limit itself is a consequence of detection in the diffraction plane, and it applies to 4D-STEM imaging generally, including bright-field, dark-field, differential-phase-contrast \cite{Shibata-etal2012}, center-of-mass \cite{Muller-etal2014,Yucelen-etal2018}, matched-illumination \cite{Ophus-etal2016,Yang-etal2016}, symmetry-based \cite{KrajnakEtheridge2020}, and ptychographic \cite{Nellist-etal1995,Putkunz-etal2012,Jiang-etal2018,Song-etal2019,Zhou-etal2020,Schloz-etal2020,ChenMuller-etal2020,Li-etal2022} imaging. 

We compare the dose efficiency of 4D-STEM with phase-contrast transmission electron microscopy (TEM), the standard (direct, real space) imaging modality for biological materials and whose collection efficiency is similar to 4D-STEM. Under the Zernike phase condition, phase-contrast TEM provides the greatest sensitivity, in that it enables the quantum limit for all spatial frequencies admitted by the optics. While 4D-STEM generally cannot attain the quantum limit, it can yield information on spatial frequencies well beyond those accessible by phase-contrast TEM.

\section{Background}

\begin{figure}
\includegraphics[width=1\columnwidth]{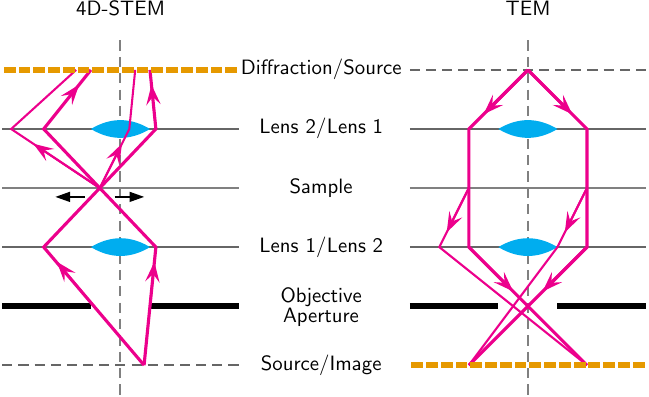}
\caption{\label{fig: STEM imaging setup} Electron-optical geometries for 4D-STEM (left) and phase-contrast TEM (right). These reciprocally-related geometries have similar collection efficiencies, and we assume an equivalent degree of aberration control up to the field angles admitted by the objective aperture. In reality, the STEM probe is scanned using deflector coils (omitted).}
\end{figure}

We will consider the 4D-STEM and TEM optical setups in Fig.~\ref{fig: STEM imaging setup}, where beams of $\sim$100~keV electrons pass through an electron-transparent sample. In 4D-STEM, a pixellated detector captures the diffracted intensity distribution for each beam position, and the resulting four-dimensional dataset is used to reconstruct an image. In TEM, we assume fixed parallel illumination at normal incidence, and a pixellated detector in the image plane captures the image directly.

We assume that the experimental goal is to estimate, simultaneously, a set of $P$ real parameters $\lambda_1,\dots,\lambda_P$. In the formalism of multiparameter quantum estimation theory, the parameters do not require a representation in terms of an Hermitian operator \cite{Braunstein-etal1996}, and it need only be the case that the detected quantum state depends (effectively) continuously on them. We also assume that the initial (incident) quantum state does not depend on the parameters. The latter assumption enables considerable simplification of the theory, though it does exclude the potentially very interesting possibility of tuning the initial state to the parameters.

In this work, we take the parameters to be the moduli and phases of the Fourier coefficients of the sample's projected electrostatic potential $\V$. For materials structure determination, the phases of the Fourier coefficients are usually of particular importance since their values often dominate the positions of features in the sample. However, in the results presented below, we shall mostly be able to treat the Fourier moduli and phases on equal footing. The moduli and phases of the Fourier coefficients form our set of $P$ real parameters $\lambda_1,\dots,\lambda_P$. We assume that all other parameters, such as those characterizing the optics, are already known with sufficient accuracy. 

\section{Quantum and Classical Fisher information}

The quantum Fisher information matrix (QFIM), a $P\times P$ matrix denoted $J$, is a key quantity in quantum estimation theory \cite{Liu-etal2020}. $J$ is a quantum analogue of the usual (classical) Fisher information matrix (CFIM), a $P\times P$ matrix denoted $I$. In the regime of asymptotic statistics, $I$ and $J$ are related to the attainable variance in the (unbiased) estimation of a parameter $\lambda_\mu$ via the inequality chain \cite{BraunsteinCaves1994,Braunstein-etal1996}
\beq\label{eq:inequality chain} \var{\lambda_\mu} \ge {\Fc}^{-1}_{\mu\mu} \ge {\Fq}^{-1}_{\mu\mu}. \eeq
$I^{-1}_{\mu\mu}$ is the $\mu$th diagonal element of the inverse matrix $I^{-1}$, and analogously for $J^{-1}_{\mu\mu}$. Thus, $I^{-1}_{\mu\mu}$ gives the Cramer-Rao lower bound which applies to any (unbiased) estimator, and $J^{-1}_{\mu\mu}$ provides a lower bound for $I^{-1}_{\mu\mu}$. If \emph{both} \emph{equalities} are obtained for all parameters, i.e., $\var{\lambda_\mu} = {\Fq}^{-1}_{\mu\mu}$ $\forall\mu$, then the \emph{simultaneous quantum limit} is achieved. The first equality can be achieved using a suitable estimator, such as a maximum-likelihood estimator. However, the second equality can be achieved only under optimum experimental conditions. 

For a pure quantum state $\ket{\psi}$, $J$ can be defined as \cite{Liu-etal2020}
\beq \label{eq: qfim} J_{\mu\nu}  \equiv 4N \mathrm{Re}\,\matel{\psi_\mu}{\Q}{\psi_\nu}, \eeq
where
\beq \ket{\psi_\mu} \equiv \del \ket{\psi}/\del\lambda_\mu,\eeq
$\Q$ is the projector onto the orthogonal complement of $\ket{\psi}$, given by
\beq \Q \equiv 1 - \ketbra{\psi}{\psi},\eeq
and $N$ is the number of independent repetitions of the experiment. In our context, $N$ is the number of beam electrons used.

To define the CFIM elements $I_{\mu\nu}$, we assume that the detection of $\ket{\psi}$ is described by a projection-valued measure (PVM) specified by a complete set of projectors $\set{\ketbra{\xi}{\xi}}$. Each projector corresponds to a possible experimental outcome with probability $p(\xi) = |\braket{\xi}{\psi}|^2$. In this case, $I_{\mu\nu}$ can be written in the form
\beq\label{eq: cfim} I_{\mu\nu} = 4N\sum_\xi \frac{\mathrm{Re}\{\braket{\psi_\mu}{\xi}\braket{\xi}{\psi}\}\mathrm{Re}\{\braket{\psi}{\xi}\braket{\xi}{\psi_\nu}\}}{\braket{\xi}{\psi}\braket{\psi}{\xi}}.\eeq

It is important to appreciate that, while $J$ involves the state $\ket{\psi}$, it does \emph{not} involve the process of detection. By contrast, $I$ \emph{does} also depend on the specifics of the detection process as represented by the PVM. Loosely, we can think of $J$ and $I$ as the ``potential" and ``actual" information, respectively. An experiment enables the quantum limit if $I=J$, which is possible if, and only if \cite{Matsumoto2002,BaumgratzDatta2015,Pezze-etal2017,Liu-etal2020,BelliardoGiovannetti2021},
\begin{subequations}
\label{eq: conditions 1 and 2}
\begin{align} \label{eq: condition 1} \matel{\psi}{\comm{\H_\mu}{\H_\nu}}{\psi} = 0  \quad\text{$\forall$ $\mu$ and $\nu$},\\
\label{eq: condition 2} \braket{\psi}{\xi}\matel{\xi}{\Q}{\psi_\mu} \in\mathbb{R}  \quad\text{$\forall$ $\xi$ and $\mu$}.
\end{align}
\end{subequations}
In the commutativity condition \eqref{eq: condition 1}, $\H_\mu$ is an Hermitian generator for $\lambda_\mu$, and the generators must commute on the Hilbert space of $\ket{\psi}$. The reality condition \eqref{eq: condition 2} has a simple geometric interpretation, namely, that the rays in the Argand plane representing $\braket{\xi}{\psi}$ and $\matel{\xi}{\Q}{\psi_\mu}$ are parallel, corresponding to maximally strong interference.

In what follows, unless otherwise stated, we adopt the weak phase-object approximation (WPOA) whereby the projected potential $\V$ is regarded as small compared to unity, and expressions are retained to \emph{leading} order (not necessarily first order) in $\V$. In the case of 4D-STEM, retaining terms to leading order is necessary for including the dark field contributions to $I$ and $J$. While the dark-field intensity is one order in $\V$ higher than the bright-field intensity, $I$ and $J$, in fact, involve \emph{changes} in the scattered amplitudes with respect to the parameter values, as opposed to the scattered intensities. Hence, in 4D-STEM, the bright- and dark-field contributions to $I$ and $J$ are of the same order in $\V$. 

We emphasize that the general theory is not restricted to the WPOA, and we refer the reader to our previous work \cite{Dwyer2023} for expressions pertaining to scattering conditions ranging from weak to strong. On the other hand, the simplicity of the WPOA allows analytical results which build intuition and pave the way for future work. In the WPOA, condition \eqref{eq: condition 1} is always satisfied, since the $\H_\mu$'s reduce to the $\V_\mu$'s (defined in Sec.~\ref{sec: parameters}), and the latter always commute.

\section{Our parameters}\label{sec: parameters}

In coordinate space, the sample's projected electrostatic potential $\V$ can be written in the form
\beq V(x) = \sum_k V(k)e^{2\pi i k\cdot x}, \eeq
where $x$ and $k$ are two-dimensional vectors in the plane transverse to the optic axis, and the Fourier coefficients $V(k)$ obey Friedel symmetry $V(k) = \bar V(-k)$, consistent with $V(x)$ being a real-valued function (see Appendix \ref{Appendix: conventions} for further conventions). 

In this work, we choose our parameters $\set{\lambda_\mu}$ to be (a subset of) the Fourier moduli $|V(k_\mu)|$ and phases $\phi(k_\mu) \equiv \arg V(k_\mu)$. The two-dimensional spatial frequency $k_\mu$ carries a subscript $\mu$ to specify that it is associated with the parameter set. Many of our expressions below apply to both the Fourier moduli and phases, but when needed we will further specify whether $\lambda_\mu$ means $|V(k_\mu)|$ or $\phi(k_\mu)$. Owing to the Friedel symmetry, we restrict $k_\mu$ to the half space defined by, e.g., the union of regions $k_x>0$ and $k_x=0,k_y\ge0$. Note that our definition of the half space includes the zero spatial frequency $k_\mu=0$. However, we find that both the quantum and classical Fisher information on $V(k_\mu=0)$ vanishes, because it corresponds to knowledge of the overall phase of $\ket{\psi}$, which is not observable. With this understood, we find it easiest to simply exclude the case $k_\mu=0$ in the mathematical expressions in Secs.~\ref{sec: TEM} and \ref{sec: 4D-STEM} (even though some expressions would remain valid). 

In the following sections, we will need the derivatives of $\V$ with respect to each $\lambda_\mu$. These derivatives will be denoted $\V_\mu$, and they have the following Fourier representations 
\beq \label{eq: dV modulus}\begin{split} & V_\mu(k) = \begin{cases} \delta_{k,k_\mu}e^{i\phi(k_\mu)} + \delta_{k,-k_\mu}e^{-i\phi(k_\mu)}, & \text{$\lambda_\mu = |V(k_\mu)|$},\\
\delta_{k,k_\mu} i V(k_\mu) - \delta_{k,-k_\mu} i \bar V(k_\mu), & \text{$\lambda_\mu = \phi(k_\mu)$},  \end{cases}\\
 &\qquad \text{$|k_\mu|>0$}, \end{split}\eeq
where $k$ is arbitrary (and $k_\mu$ is in the half space excluding the origin).

\section{Phase-contrast TEM}\label{sec: TEM}

We let the incident state $\ket{\psi_0}$ be a plane wave at normal incidence, denoted $\ket{k_0}$ with $k_0=0$. We obtain
\beq \ket{\psi} = \A (1-i\V)\ket{k_0},\eeq
where $\A$ is the \emph{non}unitary operator
\beq \A \equiv \sum_{|k|\le K} \ket{k}e^{-2\pi i\chi(k)} \bra{k}, \eeq
$\chi(k)$ is the aberration phase shift, and $K$ is the objective aperture radius. The nonunitarity of $\A$ arises because the aperture blocks some of the scattering. 

With the above expressions, we find (Appendix~\ref{Appendix: J for TEM}) that the QFIM for phase-contrast TEM imaging is diagonal, with
\beq \label{eq: qfim tem} J_{\mu\mu} = 8N |V_\mu(k_\mu)|^2, \qquad \text{$0 < |k_\mu| \le K$}.\eeq
This expression applies to both Fourier moduli and phases, and it is independent of the aberrations. Using \eqref{eq: qfim tem} in \eqref{eq:inequality chain}, we obtain for the variances
\beq\begin{split}\label{eq: variances} \var{|V(k_\mu)|} &\ge 1/8N ,\\
\var{\phi(k_\mu)} &\ge 1/8N|V(k_\mu)|^2,\end{split}\eeq
where the equalities correspond to the quantum limit. Both of the above variances will tend to vary inversely with $N$, as expected. Also, the phase variance varies inversely with the modulus, and since the moduli tend to decrease with increasing $|k_\mu|$, the phase variance will tend to increase with increasing $|k_\mu|$, also as expected.

\begin{figure}
\includegraphics[width=1\columnwidth]{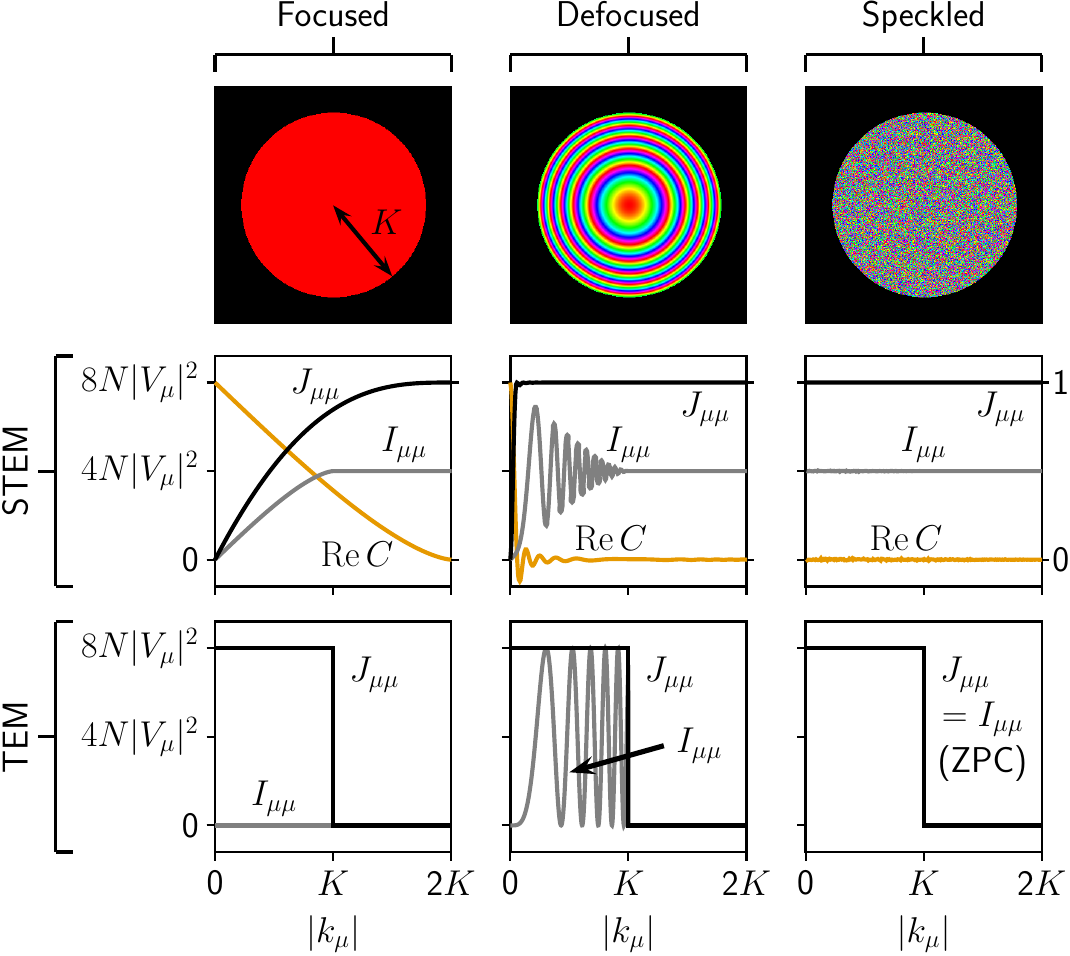}
\caption{\label{fig: results} Quantum $J_{\mu\mu}$ and classical $I_{\mu\mu}$ Fisher information on the Fourier moduli and phases obtained from 4D-STEM and phase-contrast TEM. Results are presented for diverse aberration conditions. For each aberration condition, the image shows an RGB plot of the aberration phase shift within the objective aperture, and graphs show scaled plots of $J_{\mu\mu}$ and $I_{\mu\mu}$ for spatial frequencies up to $2K$. The lower-right graph shows TEM results for the Zernike phase condition (ZPC). For STEM, the real part of the autocorrelation $C$ is also shown (refer to right axis). In all cases, and in some cases despite appearances, the values of $I_{\mu\mu}$ and $J_{\mu\mu}$ at $|k_\mu|=0$ are zero. The scaled values of $I_{\mu\mu}$, $J_{\mu\mu}$ and $C$ at $|k_\mu|=2K$ persist at higher spatial frequencies. A $100$~keV electron beam, $20$~mrad aperture semi-angle, and $-100$~nm defocus are assumed.}
\end{figure}

To determine whether phase-contrast TEM imaging permits the quantum limit, we adopt for the PVM the complete set of projectors onto coordinate space $\set{\frac{1}{M}\ketbra{x}{x}}$ ($M$ is the number of points in the discretization of coordinate space, see Appendix~\ref{Appendix: conventions}). Alternatively to considering \eqref{eq: condition 2}, we can calculate the CFIM for TEM directly (Appendix~\ref{Appendix: I for TEM}), which gives
\beq \label{eq: cfim tem} \begin{split} I_{\mu\mu} &= 4N |V_\mu(k_\mu)|^2\\
&\times\left( 1 - \cos[2\pi(2\chi(0) - \chi(k_\mu) - \chi(-k_\mu)) ] \right),\\
&\qquad\qquad\qquad\qquad\qquad\qquad\qquad 0 < |k_\mu| \le K.\end{split} \eeq 
This expression applies to both Fourier moduli and phases, and, unlike the expression for $J_{\mu\mu}$, it clearly does depend on the aberrations, as we should expect. (We note that while \eqref{eq: cfim tem} has the appearance of a contrast transfer function, we emphasize that the meaning and the scale are different.) 

In Fig.~\ref{fig: results} (bottom row) we compare $J_{\mu\mu}$ and $I_{\mu\mu}$ from TEM phase-contrast imaging for three qualitatively different aberration conditions. (Note that the plots of $J_{\mu\mu}$ and $I_{\mu\mu}$ in Fig.~\ref{fig: results} are scaled such that the dependence on the factor $8N |V_\mu(k_\mu)|^2$ is removed. In this way, the plots apply to both Fourier moduli and phases, although in the case of the phases, the natural decay of $J_{\mu\mu}$ and $I_{\mu\mu}$ with increasing $|k_\mu|$, owing to the decay of the moduli with increasing $|k_\mu|$, is masked.) For perfect focus $\chi(k)=0$, the phase-contrast TEM image contains no information on the Fourier moduli or phases, as we should expect. A defocused condition enables the quantum limit for specific spatial frequencies. A Zernike phase condition $\chi(0)=\frac{1}{4}, \chi(k\neq0)=0$ enables the quantum limit for the Fourier moduli and phases at all spatial frequencies admitted by the optics \cite{Koppell-etal2022,Dwyer2023}. An absolutely key point is that the Zernike condition makes $\ket{\psi}$ \emph{real} (up to overall phase), so that $\ket{\psi}$ entails optimal interference with greatest possible sensitivity to the parameters of $\V$. The latter statement corresponds to the satisfaction of the reality condition \eqref{eq: condition 2}.

\section{4D-STEM}\label{sec: 4D-STEM}

We regard the 4D-STEM experiment as $M$ independent quantum systems, for which the total quantum state is the tensor product
\beq \ket{\Psi} = \ket{\psi(x_1)}\otimes\cdots\otimes\ket{\psi(x_M)}, \eeq
where $\ket{\psi(x)}$ is a scattered state for which the incident beam was positioned at $x$ in the sample plane. We then use the fact that $J$ (and $I$) is additive with respect to independent systems \cite{Lu-etal2012,*TothApellaniz2014}. We also introduce the standard notation for the STEM probe wave function $\braket{k}{\psi_0(x)} =\psi_0(k)e^{-2\pi i k\cdot x}$, where
\beq\label{eq: stem wave function} \psi_0(k) = \begin{cases} |\psi_0(k)|e^{-2\pi i\chi(k)} & \text{$|k| \le K$},\\
0 & \text{otherwise}. \end{cases} \eeq

The QFIM for STEM is found (Appendix~\ref{Appendix: J for 4D-STEM}) to be diagonal, with
\beq \label{eq: qfim stem} J_{\mu\mu} = 8N |V_\mu(k_\mu)|^2(1- |C(k_\mu)|^2),\qquad \text{$|k_\mu|>0$},\eeq
where $C(k_\mu) = \sum_{k} \psi_0(k)\bar\psi_0(k+k_\mu)$ is an autocorrelation (with $C(0)=1$). Once again, this expression for $J_{\mu\mu}$ applies to both Fourier moduli and phases. However, unlike $J_{\mu\mu}$ for TEM, \eqref{eq: qfim stem} \emph{does} depend on the aberrations through $C(k_\mu)$, and $|k_\mu|$ \emph{can} be greater than $K$. Maximum quantum Fisher information is obtained when $|C(k_\mu)|$ is negligible compared to unity (see discussion), and in this case the variances obey the inequalities \eqref{eq: variances} in Sec.~\ref{sec: TEM} (with no upper bound on $|k_\mu|$). 

Notwithstanding the above remarks, we find that 4D-STEM does \emph{not} enable the quantum limit for the Fourier modulus or phase at any spatial frequency, and typically it can enable only \emph{half} of this limit. To see why, we adopt for the PVM the complete set of projectors onto Fourier space $\set{\ketbra{k}{k}}$, and we consider the reality condition \eqref{eq: condition 2}. 

\begin{figure}
\includegraphics[width=0.7\columnwidth]{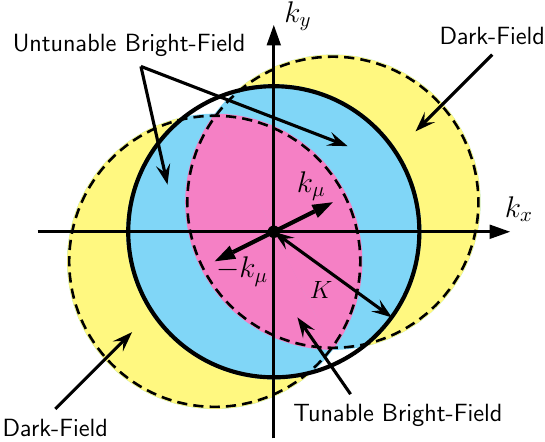}
\caption{\label{fig: bright-field} Bright- and dark-field contributions to the classical Fisher information for a spatial frequency $|k_\mu|\le K$. Solid circle represents the STEM objective aperture. Dashed circles represent the aperture displaced by $\pm k_\mu$. For $K < |k_\mu|\le 2K$ (not shown) there is no tunable region. For $|k_\mu| > 2K$ (not shown) there are only dark field contributions.}
\end{figure}

For wave vectors $k$ in the bright field, assuming that $|C(k_\mu)|\approx 0$, \eqref{eq: condition 2} becomes (Appendix~\ref{Appendix: quantum limit conditions for 4D-STEM})
\beq\label{eq: condition 2 stem wpoa}\begin{split}
 &-i \bar\psi_0(k) \psi_0(k-k_\mu) V_\mu (k_\mu) e^{+2\pi ik_\mu\cdot x} \\
 &+ (+k_\mu\rightarrow -k_\mu)\in\mathbb{R}\qquad\text{$\forall$ $x$, $|k|\le K$ and $k_\mu$}, \end{split}\eeq
where the notation in parentheses implies the preceding term with $+k_\mu$ replaced by $-k_\mu$. If $|k-k_\mu| \le K$ and $|k+k_\mu| \le K$ (the ``tunable region," see Fig.~\ref{fig: bright-field}), then the two terms in \eqref{eq: condition 2 stem wpoa} can combine to become real if the aberrations are such that $2\chi(k)-\chi(k-k_\mu)-\chi(k+k_\mu) = n +\tfrac{1}{2}$ for some integer $n$ (note that the ``tunable region" is where a bright-field detector is typically placed to generate a phase-contrast STEM image). On the other hand, if only $|k-k_\mu| \le K$ or $|k+k_\mu| \le K$ (the ``untunable region," see Fig.~\ref{fig: bright-field}), then only one of the terms in \eqref{eq: condition 2 stem wpoa} is in effect, and due to the phase factor involving $x$, that term varies continuously between purely real and purely imaginary, regardless of $k_\mu$ or $\chi$, with the effect being as though \eqref{eq: condition 2 stem wpoa} is satisfied for only \emph{half} of the beam positions. Since the untunable region typically comprises a significant portion of the bright field, condition \eqref{eq: condition 2 stem wpoa} cannot be satisfied in general. 

For $k$ in the dark field, condition \eqref{eq: condition 2} takes the form (Appendix~\ref{Appendix: quantum limit conditions for 4D-STEM})
\beq\label{eq: condition 2 stem dark field}\begin{split}
 &\sum_{k'}\bar\psi_0(k-k') \bar V(k') V_\mu (k_\mu) \psi_0(k-k_\mu)e^{-2\pi i(k'-k_\mu)\cdot x}\\
 &+ (+k_\mu\rightarrow -k_\mu) \in\mathbb{R} \qquad\text{$\forall$ $x$, $|k| > K$ and $k_\mu$}, \end{split}\eeq
where the notation implies another entire summation with $+k_\mu$ replaced by $-k_\mu$. For a given $k$, only one of these summations can be in effect, which means that we cannot balance terms as before. Also notice that \eqref{eq: condition 2 stem dark field} is \emph{non}linear in $\V$, which prevents us from obtaining general results. However, when multiple spatial frequencies $k'$ contribute to the summation, as occurs when the STEM probe convergence angle is large enough for there to be large overlap of the diffraction discs, then, similarly to the untunable bright-field, owing to the phase factors involving $x$, it is again as though \eqref{eq: condition 2 stem dark field} is satisfied for only \emph{half} of the beam positions, regardless of $k_\mu$ or $\chi$. 

There are exceptions to the behavior of the dark field just described: If the term $k'=k_\mu$ dominates the summation, as it can when the convergence angle is small and there is negligible overlap of the diffraction discs, then \eqref{eq: condition 2 stem dark field} is always satisfied for $\lambda_\mu = |V(k_\mu)|$, i.e., full modulus information, and never satisfied for $\lambda_\mu = \phi(k_\mu)$, i.e., no phase information. The situation just described is the classic ``phase problem" in parallel-beam diffraction. As the convergence angle is increased, such that the overlap increases from negligible to large, we can infer that the behavior interpolates between ``full modulus information and no phase information" and ``half modulus information and half phase information."

For definiteness, we will assume that the overlap of diffraction discs in the 4D-STEM experiment is large, such that multiple spatial frequencies contribute in \eqref{eq: condition 2 stem dark field}, and the phase information is maximized. In this case, we find (Appendix~\ref{Appendix: I for 4D-STEM}) that the \emph{complete} CFIM for 4D-STEM is approximately diagonal, with diagonal elements given by
\beq\begin{split} \label{eq: cfim stem} &I_{\mu\mu} \approx 4N|V_\mu(k_\mu)|^2 \\
&\times \big(1 - \sum_k |\psi_0(k-k_\mu)| |\psi_0(k+k_\mu)| \\
&\qquad\times \cos[2\pi(2\chi(k)-\chi(k-k_\mu)-\chi(k+k_\mu))]\big),\\
&\text{$|k_\mu|>0$.} \end{split}\eeq
This expression applies to both Fourier moduli and phases. The terms inside the summation are the tunable bright-field contributions, which can range from fully additive (for the previously-stated condition on $\chi$) to fully subtractive (e.g., when $\chi=0$).

In Fig.~\ref{fig: results} (middle row), we compare $J_{\mu\mu}$ and $I_{\mu\mu}$ from 4D-STEM for different aberration conditions. Perfect focus results in minimum $J_{\mu\mu}$ (since $|C|$ is maximized) and minimum $I_{\mu\mu}$ (since the tunable contributions are fully subtractive). A defocused condition dramatically improves $J_{\mu\mu}$ and improves $I_{\mu\mu}$ at specific spatial frequencies though not others. For an arbitrary spatial frequency $k_\mu$, a near-optimum $\chi$ is one that is ``random" on $[0,2\pi)$,  giving $J_{\mu\mu} \approx 8N |V_\mu(k_\mu)|^2$ and $I_{\mu\mu}\approx\frac{1}{2}J_{\mu\mu}$. The latter case corresponds to a delocalized speckled probe.

\section{Discussion and Conclusions}

Let us first review the meaning of the inequality chain \eqref{eq:inequality chain}. The QFIM $J$ can be regarded as the potential information in the scattered quantum state before detection, and the CFIM $I$ regarded as the actual information contained in the detected scattering. Optimum experimental conditions, as embodied by \eqref{eq: conditions 1 and 2}, result in $I=J$ (second equality in \eqref{eq:inequality chain}). Finally, extraction of all information $I$ on the parameters [first equality in \eqref{eq:inequality chain}] requires suitable estimators (e.g., maximum-likelihood). In phase-contrast TEM, Fourier analysis of the image intensity provides suitable estimators of the Fourier coefficients (under the WPOA). In 4D-STEM, the estimators comprise a computational algorithm that generate values of the Fourier coefficients from the scattering data. In this work, we do not consider the latter estimators in any detail, so our comments apply to 4D-STEM techniques generally.

In the calculations in Section \ref{sec: 4D-STEM}, we have assumed that the 4D-STEM detector captures the entire scattering distribution, and found that the CFIM is approximately half of the QFIM. Thus, 4D-STEM, with its detector positioned in the diffraction plane, precludes the quantum limit in the simultaneous estimation of the Fourier moduli and phases. The latter statement is independent of the way in which the scattering data is processed, and thus applies to any 4D-STEM technique. In fact, this conclusion applies even more broadly to similar techniques using other forms of coherent scalar radiation, such as visible light and x-rays. The CFIM will be further reduced by a less-capable detector, such as one that does not resolve any fine features in the scattering distribution, or one that does not capture the entire distribution.

In 4D-STEM, the aberration-dependence of $J_{\mu\mu}$ can reduce the quantum information for $|k_\mu| \le 2K$. For $k_\mu$ arbitrary, $J_{\mu\mu}$ is near-maximized by a ``random" $\chi$, producing a delocalized speckled probe with expected autocorrelation $\av{|C(k_\mu)|^2}\le 1/M_K$ ($M_K$ is the number of plane waves inside the aperture) and $I_{\mu\mu}\approx\frac{1}{2}J_{\mu\mu}$. This provides theoretical grounding for previous empirical observations made in the context of 4D-STEM ptychography \cite{Pelz-etal2017}, as well as light and x-ray classical-imaging settings \cite{Ventalon2005,Mudry2012,Gatti2006,Berujon2012,Morgan2012}. Low-autocorrelation sequences \cite{BorweinFerguson2005} provide scope for minor further optimization of $J_{\mu\mu}$. The $\chi$ developed in the work of Ophus et al.~\cite{Ophus-etal2016} for information transfer in the bright field is slightly less optimum than random. A defocused $\chi$ is most practical using current STEMs, though, as shown in Fig.~\ref{fig: results}, the CFIM is nonuniform across the spatial frequencies. Owing to the additivity of $I$ (and $J$), the CFIM in the defocused case can be made significantly more uniform by acquiring and processing data at multiple defoci.

Note that $\chi$ must be effectively known to extract information $I_{\mu\mu}$ from the scattering data. If additional parameters, such as those characterizing the optics, are not already known with sufficient accuracy, then they should be included in the set of parameters to be estimated. This will potentially decrease the precision achievable for the Fourier coefficients. In such cases, our results for $I_{\mu\mu}$ and $J_{\mu\mu}$ should be regarded as upper bounds.

We have compared 4D-STEM with Zernike phase-contrast TEM, which does enable the quantum limit for all spatial frequencies admitted by the objective aperture. Therefore, in principle, for the spatial frequencies that it can access, Zernike phase contrast can match the precision of 4D-STEM using about half of the electron dose. However, the realization of robust Zernike phase plate for electrons is highly nontrivial \cite{Rose2010,Glaeser2013,Majorovits-etal2007,Lentzen2008,NagayamaDanev2008,Alloyeau-etal2010,Danev-etal2014,Muller-etal2010,Schwartz-etal2019}. A phase plate is not necessary for ``conventional" 4D-STEM, though one is necessary for a speckled probe.

The results presented in the main text of this work assumed perfect coherence for simplicity. Taking into account partial spatial coherence (Appendix~\ref{Appendix: partial spatial coherence}) does change some details, but it does not change our broad conclusions: In phase-contrast TEM, partial spatial coherence reduces $J$ at very low spatial frequencies, reduces $I$ also at spatial frequencies where the aberration is varying, but a Zernike condition still enables the quantum limit. In 4D-STEM, partial spatial coherence reduces $J$ at spatial frequencies $|k_\mu|\le2K$, and reduces the elements of $I$ that refer to the phases at all spatial frequencies, but detection in the diffraction plane still precludes the quantum limit. We have not yet made a mathematical analysis of the effect of partial temporal coherence. However, we anticipate that the most significant change will be that, in phase-contrast TEM, it will not affect $J$ but it will reduce $I$ at higher spatial frequencies, meaning that a Zernike condition will no longer enable the quantum limit at those higher spatial frequencies (anticipation of this effect was, in fact, our reason for introducing an objective aperture in the TEM setup).

While it does not enable the quantum limit at any spatial frequency, 4D-STEM is able to provide information for spatial frequencies well beyond those accessible by phase-contrast imaging (for the same degree of aberration control). 4D-STEM also provides more flexibility, since a broad range of image types can be derived. Thus, regarding a choice between the two forms of imaging, based on the present analysis, Zernike phase-contrast TEM should provide greatest sensitivity for resolutions up to about 1~\AA, whereas 4D-STEM should be used when larger datasets can be tolerated, flexibility is beneficial, and information is desired at deep-sub-\AA\ resolutions. We also mention that ptychographical techniques based on 4D-STEM can enable estimates of the sample's electrostatic potential under strong scattering conditions \cite{Maiden-etal2012,Brown-etal2018,Chen-etal2021}, which is another significant advantage.

Lastly, we remark that the trade off between dose efficiency and spatial resolution in TEM and STEM has been discussed for decades \cite{Thomson1973,Rose1974,Rose1975}. However, what is different about the formalism used here is its generality. This is apparent from the fact that our analyses required no assumptions about how the experimental data is processed. Moreover, the consideration of electron dose is an integral part of the formalism, rather than having to be inferred from additional calculations. Finally, the present formalism readily exhibits the ultimate limits of precision as allowed by the laws of quantum mechanics, and it allows some deeper, significant insights. For example, in the case of Zernike phase-contrast imaging, that the optical setup renders the detected quantum state real is the deeper reason why the quantum limit can be attained. For 4D-STEM, with its indirect image formation via Fourier space, the detected quantum state is inherently complex, and the quantum limit is precluded.

\acknowledgements

We acknowledge fruitful discussions with Tim Petersen (Monash University) and Daniel Stroppa (Dectris Ltd.).

\begin{widetext}

\appendix
\section{Conventions}\label{Appendix: conventions}

We use the following conventions
\beq \mathbbm{1} = \sum_k \ketbra{k}{k}=\frac{1}{M} \sum_x \ketbra{x}{x},\eeq
where $M$ is the number of points in the discretized 2D coordinate or Fourier space. Also $\braket{x}{x'}=M\delta_{xx'}$, $\braket{k}{k'}=\delta_{kk'}$, and $\braket{x}{k}=e^{2\pi ik\cdot x}$. These produce for the discrete Fourier transform of $\psi(x) = \braket{x}{\psi}$ and its inverse
\beq \psi(k) = \braket{k}{\psi} = \frac{1}{M}\sum_x \braket{k}{x}\braket{x}{\psi} = \frac{1}{M}\sum_x e^{-2\pi ik\cdot x}\psi(x), \eeq
and
\beq \psi(x) = \braket{x}{\psi} = \sum_k \braket{x}{k}\braket{k}{\psi} = \sum_k e^{2\pi ik\cdot x}\psi(k).\eeq
Normalization of the wave functions is given by
\beq 1 = \braket{\psi}{\psi} = \sum_k |\psi(k)|^2 = \frac{1}{M} \sum_x|\psi(x)|^2.\eeq 
The Fourier and coordinate representations of $\V$ are given by
\beq\begin{split} \V &= \sum_{kk'} \ketbra{k}{k}\V \ketbra{k'}{k'} = \sum_{kk'} \ket{k}V(k-k') \bra{k'}\\
&= \frac{1}{M^2}\sum_{xx'} \ketbra{x}{x}\V \ketbra{x'}{x'} = \frac{1}{M^2} \sum_{xx'} \ket{x}V(x) M \delta_{xx'} \bra{x'} = \frac{1}{M} \sum_{x} \ket{x}V(x) \bra{x}. \end{split} \eeq
The discrete Fourier transform and inverse transform of $\V$ are given by
\beq V(k-k') = \matel{k}{\V}{k'} = \frac{1}{M^2}\sum_{xx'}\braket{k}{x}\matel{x}{\V}{x'}\braket{x'}{k'} = \frac{1}{M}\sum_{x}e^{-2\pi i (k-k')\cdot x} V(x),  \eeq
and
\beq V(x) = \frac{1}{M}\matel{x}{\V}{x} = \frac{1}{M}\sum_{kk'}\braket{x}{k+k'}\matel{k+k'}{\V}{k'}\braket{k'}{x} = \sum_{k}e^{2\pi ik\cdot x} V(k). \eeq
By $\V$, we mean the projected electrostatic interaction energy times $1/\hbar v$, where $v$ is the beam electron speed, and $V(x)$ is negative for a beam electron interacting with an atom. Analogous expressions hold for $\V_\mu$.

\section{Calculation of $J_{\mu\nu}$ for phase-contrast TEM}\label{Appendix: J for TEM}

For phase-contrast TEM, the detected state in the WPOA is given by
\beq \ket{\psi} = \A(1-i\V)\ket{k_0}.\eeq
To leading order in $\V$, we obtain for the QFIM
\beq J_{\mu\nu} = 4N \mathrm{Re}\,\matel{\psi_\mu}{\Q}{\psi_\nu} = 4N\mathrm{Re}\, \matel{k_0}{\V_\mu \adj\A (1-\ketbra{k_0}{k_0})\A\V_\nu}{k_0}. \eeq
Using the expansion
\beq \A = \begin{cases} \sum_k \ket{k}e^{- 2\pi i\chi(k)}\bra{k}, & \text{$|k|\le K$},\\
0 & \text{otherwise}, \end{cases}\eeq
we get
\beq\begin{split} J_{\mu\nu} &= 4N \left(\sum_{|k|\le K} \mathrm{Re}\,\matel{k_0}{\V_\mu}{k}\matel{k}{\V_\nu}{k_0} - \matel{k_0}{\V_\mu}{k_0}\matel{k_0}{\V_\nu}{k_0} \right)\\
&= 4N \left(\sum_{|k|\le K} \mathrm{Re}\,\bar V_\mu(k) V_\nu(k) - V_\mu(0) V_\nu(0) \right).
\end{split} \eeq
This vanishes unless $k_\mu=k_\nu$ and $0<|k_\mu|\le K$, in which case we obtain
\beq J_{\mu\nu} = 4N \left(\bar V_\mu(k_\mu) V_\nu(k_\mu) + V_\mu(k_\mu) \bar V_\nu(k_\mu)\right), \qquad\text{$0<|k_\mu|\le K$}. \eeq 
This also vanishes unless $\mu$ and $\nu$ refer to the same modulus or same phase, that is, $J_{\mu\nu}$ is diagonal. The diagonal elements are given by
\beq J_{\mu\mu} = 8N |V_\mu(k_\mu)|^2, \qquad\text{$0<|k_\mu|\le K$}, \eeq
which is Eq.~\eqref{eq: qfim tem}.

\section{Calculation of $I_{\mu\nu}$ for phase-contrast TEM}\label{Appendix: I for TEM}

We appropriately choose as the PVM the projectors onto coordinate space $\set{\frac{1}{M}\ketbra{x}{x}}$. The CFIM becomes, to leading order in $\V$,
\beq\begin{split} I_{\mu\nu} &= N\sum_x\frac{p_\mu(x)p_\nu(x)}{p(x)} \\
&= \frac{4N}{M}\sum_x\frac{\mathrm{Re}\{\matel{k_0}{\V_\mu\adj\A(i)}{x}\matel{x}{\A}{k_0}\}\mathrm{Re}\{\matel{k_0}{\adj\A}{x}\matel{x}{(-i)\A\V_\nu}{k_0}\}}{\matel{x}{\A}{k_0}\matel{k_0}{\adj\A}{x}}.\end{split}\eeq
Using the expansion of $\A$ given above, we obtain, for $|k_\nu|\le K$,
\beq\begin{split} \matel{k_0}{\adj\A}{x} \matel{x}{\A\V_\nu}{k_0} &= e^{2\pi i\chi(0) } \left(e^{2\pi i k_\nu\cdot x - 2\pi i\chi(k_\nu)} V_\nu(k_\nu) + e^{-2\pi i k_\nu\cdot x - 2\pi i\chi(-k_\nu)}  \bar V_\nu(k_\nu) \right) \\
&=  2 |V_\nu(k_\nu)| e^{2\pi i\chi(0)  -\pi i\chi(k_\nu) - \pi i\chi(-k_\nu)  }  \cos[2\pi  k_\nu\cdot x - \pi \chi(k_\nu) +\pi \chi(-k_\nu) + \phi_\nu(k_\nu) ].  \end{split}\eeq
The relevant real part is
\beq\begin{split} \mathrm{Re}\{\matel{k_0}{\adj\A}{x}\matel{x}{(-i)\A\V_\nu}{k_0}\} &= 2 |V_\nu(k_\nu)| \sin[2\pi \chi(0)  -\pi \chi(k_\nu) - \pi \chi(-k_\nu) ]\\
& \qquad\times  \cos[2\pi  k_\nu\cdot x - \pi \chi(k_\nu) +\pi \chi(-k_\nu) + \phi_\nu(k_\nu) ]. \end{split}\eeq
Multiplying by the analogous factor for $\mu$, and summing over $x$, we obtain that a nonzero result demands $k_\mu=k_\nu$, and then further that $\mu=\nu$, that is, $I_{\mu\nu}$ is diagonal. The diagonal elements can be cast into the form
\beq I_{\mu\mu} = 4N |V_\mu(k_\mu)|^2\left( 1 - \cos[2\pi(2\chi(0) - \chi(k_\mu) - \chi(-k_\mu))] \right),\eeq
where $|k_\mu| \le K$. This is Eq.~\eqref{eq: cfim tem}.

\section{Calculation of $J_{\mu\nu}$ for 4D-STEM}\label{Appendix: J for 4D-STEM}

We regard the STEM experiment as consisting of $M$ independent quantum systems, one system for each position of the electron beam:
\beq \ket{\Psi} = \ket{\psi(x_1)}\otimes\cdots\otimes\ket{\psi(x_M)}, \eeq
where $\ket{\psi(x)}$ is a pure scattered state for which the incident beam was positioned at $x$ in the sample plane, and $\otimes$ denotes a tensor product. Since $J$ (and $I$) is additive with respect to independent systems, we obtain
\beq J_{\mu\nu} = \frac{4N}{M}\sum_x \mathrm{Re}\,\matel{\psi_\mu(x)}{\Q(x)}{\psi_\nu(x)}, \eeq
where $\Q(x)\equiv 1-\ketbra{\psi(x)}{\psi(x)}$, $\ket{\psi_\mu(x)}\equiv\del\ket{\psi(x)}/\del\lambda_\mu$, and $M$ is the number of ``pixels" in a discretization of the two-dimensional space. With this normalization, $N$ corresponds, as in our analysis of TEM, to the total number of electrons.

Using the POA (not WPOA), we obtain
\beq\begin{split} J_{\mu\nu} &= \frac{4N}{M} \sum_x \mathrm{Re}\, \matel{\psi_0(x)}{e^{+i\V}\V_\mu (1-e^{-i\V}\ketbra{\psi_0(x)}{\psi_0(x)}e^{+i\V})\V_\nu e^{-i\V}}{\psi_0(x)}\\
&= \frac{4N}{M} \sum_x \mathrm{Re}\, \matel{\psi_0(x)}{\V_\mu (1-\ketbra{\psi_0(x)}{\psi_0(x)})\V_\nu}{\psi_0(x)}.\end{split}\eeq
For the first term (containing the identity), we obtain
\beq\begin{split} &\frac{4N}{M} \mathrm{Re}\sum_{x,k,k',k''}  \bar \psi_0(k-k') e^{-2\pi ik'\cdot x} \bar V_\mu(k') V_\nu(k'')\psi_0(k-k'')e^{2\pi ik''\cdot x}\\
&\quad = 4N \mathrm{Re}\sum_{k,k'} \bar \psi_0(k-k') \bar V_\mu(k') V_\nu(k')\psi_0(k-k')\\
&\quad = 4N \sum_{k,k'} |\psi_0(k-k')|^2 \mathrm{Re}\,\bar V_\mu(k') V_\nu(k') \\
&\quad = 4N \left(\bar V_\mu(k_\mu) V_\nu(k_\mu) + V_\mu(k_\mu) \bar V_\nu(k_\mu)\right)  ,  \end{split}\eeq
where $k_\mu$ is in the half space (defined by, e.g., $k_x>0$), and we have used $\sum_{k} |\psi_0(k)|^2=1$. From the forms of $V_\mu$ given in the main text, $\mu$ and $\nu$ must both refer to the modulus, or both refer to the phase, otherwise the expression in the last line vanishes. Hence the first term in $J_{\mu\nu}$ equals $8N|V_\mu(k_\mu)|^2\delta_{\mu\nu}$.

The second term in $J_{\mu\nu}$ is
\beq\begin{split} &-\frac{4N}{M} \mathrm{Re}\sum_{x,k,k',k'',k'''}  \bar \psi_0(k) e^{+2\pi ik\cdot x} V_\mu(k-k') \psi_0(k')e^{-2\pi ik'\cdot x} \bar\psi_0(k'') e^{+2\pi ik''\cdot x} V_\nu(k''-k''')\psi_0(k''')e^{-2\pi ik'''\cdot x}\\
&\quad =-4N \mathrm{Re}\sum_{k,k',k''}  \bar \psi_0(k) V_\mu(k-k') \psi_0(k') \bar\psi_0(k'') \bar V_\nu(k-k')\psi_0(k-k'+k'')\\
&\quad =-4N \mathrm{Re}\sum_{k,k',k''}  \bar \psi_0(k) V_\mu(k') \psi_0(k-k') \bar\psi_0(k'') \bar V_\nu(k')\psi_0(k'+k'')\\
&\quad =-4N \left(\bar V_\mu(k_\mu) V_\nu(k_\mu)+V_\mu(k_\mu)\bar V_\nu(k_\mu)\right) \Big| \sum_{k}  \bar \psi_0(k)  \psi_0(k-k_\mu)\Big|^2,
\end{split}\eeq
where, once again, the last line is nonzero only when $\mu=\nu$. Putting the two terms together, we have, for the diagonal elements
\beq\begin{split} J_{\mu\mu} &= 8N |V_\mu(k_\mu)|^2\left(1-\Big| \sum_{k}  \bar \psi_0(k)  \psi_0(k-k_\mu)\Big|^2\right)\\
&= 8N |V_\mu(k_\mu)|^2\left(1-| C(k_\mu)|^2\right),\end{split} \eeq
which is Eq.~\eqref{eq: qfim stem}. $J_{\mu\mu}$ vanishes for $k_\mu=0$ (as it does for phase-contrast TEM). If we regard $k_\mu$ as nonzero but otherwise arbitrary, then $J_{\mu\mu}$ is maximized by a single plane wave. If we further stipulate a finite aperture size $K$, then $J_{\mu\mu}$ is near-maximized by ``random" aberrations, corresponding to a delocalized speckled probe. 

We also supply the following derivation using a coordinate representation. In this space, the derivatives of the potential have the forms
\beq   V_\mu(x) = \begin{cases}  2\cos[2\pi k_\mu\cdot x + \phi(k_\mu)] & \text{for $\lambda_\mu= |V(k_\mu)|$},  \\
2|V(k_\mu)| \sin[2\pi k_\mu\cdot x + \phi(k_\mu)] & \text{for $\lambda_\mu= \arg V(k_\mu)$}. \end{cases} \eeq
In light of the above, we can set $\mu=\nu$ at the outset, and obtain
\beq\begin{split} J_{\mu\mu} &= \frac{4N}{M} \sum_x \mathrm{Re}\, \matel{\psi_\mu(x)}{e^{+i\V}\V_\mu(1-e^{-i\V}\ketbra{\psi_0(x)}{\psi_0(x)}e^{+i\V}) \V_\mu e^{-i\V} }{\psi_0(x)}\\
&= \frac{4N}{M^2} \sum_{x,x'} V^2_\mu(x') |\psi_0(x'-x)|^2 - \frac{4N}{M^3} \sum_x \left(\sum_{x'} V_\mu(x') |\psi_0(x'-x)|^2 \right)^2\\
&= \frac{4N}{M} \sum_{x'} V^2_\mu(x') - \frac{4N}{M} \sum_x \left(\frac{1}{M} \sum_{x'}   V_\mu(x') [|\psi_0(x'-x)|^2-1] \right)^2.
\end{split}\eeq
The second summation in the last line is a sum of squares. Therefore, if the spatial frequency $k_\mu$ of $V_\mu(x)$ is nonzero but otherwise arbitrary, then $J_{\mu\mu}$ is maximized by a STEM probe whose intensity in coordinate space has minimal correlation with any such $V_\mu(x)$. Apart from a plane wave (which has zero correlation with $V_\mu(x)$ so that the entire summation in question vanishes), for a finite aperture, a delocalized speckled intensity distribution has near-minimal correlation and will near-maximize $J_{\mu\mu}$.

\section{Quantum-limit conditions for 4D-STEM}\label{Appendix: quantum limit conditions for 4D-STEM}

Starting with the conditions \eqref{eq: conditions 1 and 2}, we incorporate the beam position, and we appropriately adopt for the PVM the complete set of projectors onto Fourier space $\set{\ketbra{k}{k}}$, to obtain
\begin{subequations}
\label{eq: conditions 1 and 2 stem}
\begin{align} \label{eq: condition 1 stem} \matel{\psi(x)}{\comm{\H_\mu}{\H_\nu}}{\psi(x)} = 0  \quad\text{$\forall$ $x$, $k_\mu$ and $k_\nu$},\\
\label{eq: condition 2 stem} \braket{\psi(x)}{k}\matel{k}{\Q(x)}{\psi_\mu(x)} \in \mathbb{R} \quad\text{$\forall$ $x$, $k$ and $k_\mu$}.
\end{align}
\end{subequations}
Under the WPOA, $\H_\mu=\V_\mu$, so that the commutativity condition \eqref{eq: condition 1 stem} is always satisfied (the same holds under the POA).

\subsection{Reality condition for the bright field}

For a wave vector $k$ in the bright field, the reality condition \eqref{eq: condition 2 stem} becomes, to leading order in $\V$,
\beq\begin{split} \braket{\psi(x)}{k}\matel{k}{\Q(x)}{\psi_\mu(x)} &= -i \braket{\psi_0(x)}{k}\matel{k}{(1-\ketbra{\psi_0(x)}{\psi_0(x)})\V_\mu}{\psi_0(x)} \\
&= -i \bar\psi_0(k)e^{2\pi i k\cdot x}\sum_{k',k''}[\delta_{k,k'} - \psi_0(k)\bar\psi_0(k')e^{-2\pi i(k-k')\cdot x}]V_\mu(k'-k'')\psi_0(k'')e^{-2\pi i k''\cdot x}\\
&= -i \bar\psi_0(k)\psi_0(k-k_\mu)V_\mu(k_\mu)e^{+2\pi i k_\mu\cdot x} + (+k_\mu\rightarrow-k_\mu)\\
&\qquad\qquad+i |\psi_0(k)|^2 \sum_{k'} \bar\psi_0(k')\psi_0(k'-k_\mu) V_\mu(k_\mu) e^{+2\pi ik_\mu\cdot x} + (+k_\mu\rightarrow-k_\mu)\\ 
&= -i \bar\psi_0(k)\psi_0(k-k_\mu)V_\mu(k_\mu)e^{+2\pi i k_\mu\cdot x} + (+k_\mu\rightarrow-k_\mu)\\
&\qquad\qquad +\frac{i}{M_K} C(k_\mu) V_\mu(k_\mu) e^{+2\pi ik_\mu\cdot x} + (+k_\mu\rightarrow-k_\mu) \in \mathbb{R} \quad\text{$\forall$ $x$, $k$ and $k_\mu$}, \end{split}\eeq
where $M_K = 1/|\psi_0(k)|^2$ for $|k|\le K$. $M_K$ is just the number of wave vectors inside the STEM objective aperture. Recall that we must have $|C(k_\mu)|^2\ll1$, otherwise the QFIM is significantly diminished compared with phase-contrast TEM. A diminished QFIM in STEM is achieved by using, e.g., a highly defocused probe or, better, a speckled probe, in which case $\av{|C(k_\mu)|^2}\le 1/M_K\ll 1$. We assume such a relevant case. Hence, in the last line above, we can neglect the term containing $C(k_\mu)$ to obtain
\beq -i \bar\psi_0(k)\psi_0(k-k_\mu)V_\mu(k_\mu)e^{+2\pi i k_\mu\cdot x} + (+k_\mu\rightarrow-k_\mu) \in \mathbb{R} \quad\text{$\forall$ $x$, $k$ and $k_\mu$},\eeq
which is the bright-field reality condition \eqref{eq: condition 2 stem wpoa}.

\subsection{Reality condition for the dark field}

For $k$ in the dark field, to leading order in $\V$, the projection operator $\Q(x)$ can be replaced with the identity, and the condition \eqref{eq: condition 2 stem} becomes
\beq\begin{split} \braket{\psi(x)}{k}\matel{k}{\Q(x)}{\psi_\mu(x)} &= \matel{\psi_0(x)}{\V}{k}\matel{k}{\V_\mu}{\psi_0(x)}\\
& = \sum_{k',k''} \bar\psi_0(k') V(k'-k) V_\mu (k-k'') \psi_0(k'')e^{2\pi i(k'-k'')\cdot x}\\
& = \sum_{k'}\bar\psi_0(k-k') \bar V(k') V_\mu (k_\mu) \psi_0(k-k_\mu)e^{-2\pi i(k'-k_\mu)\cdot x}
 + (+k_\mu\rightarrow -k_\mu) \in\mathbb{R}\\ &\qquad\qquad\qquad\qquad\qquad\qquad\qquad\qquad\qquad\qquad\text{$\forall$ $x$, $|k| > K$ and $k_\mu$}, \end{split}\eeq
which is the reality condition \eqref{eq: condition 2 stem dark field}.

\section{Calculation of $I_{\mu\nu}$ for 4D-STEM}\label{Appendix: I for 4D-STEM}

Using the property of additivity, it is straightforward to incorporate the beam position $x$ into the definition of the CFIM $I$:
\beq\begin{split}  &I_{\mu\nu} = \frac{N}{M}\sum_{k,x} p(k,x) (\del_\mu\ln p(k,x))(\del_\nu\ln p(k,x))\\
&=\frac{4N}{M}\sum_{k,x} \frac{\mathrm{Re}\{\braket{\psi_\mu(x)}{k}\braket{k}{\psi(x)}\}\mathrm{Re}\{\braket{\psi(x)}{k}\braket{k}{\psi_\nu(x)}\}}{\braket{k}{\psi(x)}\braket{\psi(x)}{k}}, \end{split}\eeq
where $p(k,x)=(1/M) |\braket{k}{\psi(x)}|^2$. 

\subsection{Bright-field contribution}

For the bright-field, we stipulate that $k$ lies inside the (image of the) probe-forming aperture, that is, $|k|\le K$. In the WPOA, we obtain, to leading order in $\V$,
\beq  I_{\mu\nu}^\mathrm{BF} = \frac{4N}{M}\sum_{|k|\le K,x} \frac{\mathrm{Re}\{\matel{\psi_0(x)}{i\V_\mu}{k}\braket{k}{\psi_0(x)}\}\mathrm{Re}\{\braket{\psi_0(x)}{k}\matel{k}{(-i)\V_\nu}{\psi_0(x)}\}}{\braket{k}{\psi_0(x)}\braket{\psi_0(x)}{k}}.\eeq
For the factor containing $\nu$, we obtain
\beq\begin{split} \frac{\mathrm{Re}\{\braket{\psi_0(x)}{k}\matel{k}{(-i)\V_\nu}{\psi_0(x)}\}}{|\braket{\psi_0(x)}{k}|} &=  |V_\nu(k_\nu)| |\psi_0(k-k_\nu)| \sin[ 2\pi (\chi(k)-\chi(k-k_\nu)) + 2\pi k_\nu\cdot x + \phi_\nu(k_\nu)] \\
& + |V_\nu(k_\nu)| |\psi_0(k+k_\nu)| \sin[ 2\pi (\chi(k)-\chi(k+k_\nu)) - 2\pi k_\nu\cdot x - \phi_\nu(k_\nu)] .\end{split} \eeq
A similar result is obtained for the factor containing $\mu$, and so the CFIM consists of four terms ``$+k_\mu,+k_\nu$," ``$-k_\mu,+k_\nu$," ``$+k_\mu,-k_\nu$" and ``$-k_\mu,-k_\nu$." Only the sine functions depend on the probe position $x$, and we can perform the summation over $x$ using the generic expression
\beq \frac{1}{M}\sum_x \sin[2\pi a + 2\pi k_\mu\cdot x] \sin[2\pi b + 2\pi k_\nu\cdot x] = \frac{1}{2}\delta_{k_\mu,k_\nu}\cos[2\pi(a-b)] - \frac{1}{2}\delta_{k_\mu,-k_\nu}\cos[2\pi(a+b)].\eeq
Using this expression, after some algebra, we obtain a non-zero result only for the diagonal terms
\beq\begin{split} \label{eq: cfim stem appendix} I_{\mu\mu}^\mathrm{BF} &= 4N |V_\mu(k_\mu)|^2 \sum_{|k|\le K} \left( |\psi_0(k-k_\mu)|^2 
-|\psi_0(k-k_\mu)| |\psi_0(k+k_\mu)| \cos[2\pi(2\chi(k)-\chi(k-k_\mu)-\chi(k+k_\mu))]\right).   \end{split}\eeq

\subsection{Dark-field contribution}

For the dark field, $k$ lies outside of the (image of the) probe-forming aperture, that is, $|k|> K$. To leading order in $\V$, we obtain
\beq I_{\mu\nu}^\mathrm{DF} = \frac{4N}{M}\sum_{|k|> K,x} \frac{\mathrm{Re}\{\matel{\psi_0(x)}{\V_\mu}{k}\matel{k}{\V}{\psi_0(x)}\}\mathrm{Re}\{\matel{\psi_0(x)}{\V}{k}\matel{k}{\V_\nu}{\psi_0(x)}\}}{\matel{k}{\V}{\psi_0(x)}\matel{\psi_0(x)}{\V}{k}}.\eeq
The factor $|\matel{k}{\V}{\psi_0(x)}|^2$ in the denominator cancels with the factors in the numerator, so that this expression is second order in $\V$ just like the bright field contribution. Writing each of the matrix elements $\matel{a}{b}{c}$ in the above expression in terms of its modulus $|\matel{a}{b}{c}|$ and phase $\arg \matel{a}{b}{c}$, we can obtain after some algebra
\beq\label{eq: cfim dark field} \begin{split} I_{\mu\nu}^\mathrm{DF} &= \frac{2N |V_\mu(k_\mu)||V_\nu(k_\nu)|}{M}\sum_{|k|> K,x} |\psi_0(k-k_\mu)||\psi_0(k-k_\nu)|(\delta_{\mu\nu}+\cos[\varphi_\mu(k,x)+\varphi_\nu(k,x) - 2\varphi(k,x)])\\
&+(+k_\mu\rightarrow -k_\mu,+k_\nu\rightarrow -k_\nu), \end{split}\eeq
where 
\beq \varphi_\mu(k,x) = \arg \matel{k}{\V_\mu}{\psi_0(x)} + 2\pi k\cdot x = 2\pi k_\mu\cdot x - 2\pi\chi(k-k_\mu) +\phi_\mu(k_\mu), \eeq
with an analogous expression for $\arg_\nu(k,x)$, and
\beq \varphi(k,x) = \arg\matel{k}{\V}{\psi_0(x)} + 2\pi k\cdot x = \arg \sum_{k'} V(k')\psi_0(k-k')e^{+2\pi i k'\cdot x}. \eeq
Expression \eqref{eq: cfim dark field} contains two parts, one featuring $+k_\mu,+k_\nu$ (as written out explicitly) and the other featuring $-k_\mu,-k_\nu$ (as indicated by the shorthand notation). For a given $k$, only one of those parts can be nonzero, but the summation over $k$ means that both parts always contribute. Notice that the presence of the cosine terms means that $I_{\mu\nu}^\mathrm{DF}$ is \emph{not} diagonal. Also notice that $\varphi(k,x)$ depends explicitly on the values of the Fourier coefficients participating in the summation over $k'$, which makes further simplifications of \eqref{eq: cfim dark field} difficult. However, as we will see below, the generic behavior is that the cosine terms tend to cancel out. If we make the approximation to omit the cosine terms entirely, then $I_{\mu\nu}^\mathrm{DF}$ is diagonal, with the diagonal elements taking the very simple form
\beq I_{\mu\mu}^\mathrm{DF} \approx 4N |V_\mu(k_\mu)|^2 \sum_{|k|> K} |\psi_0(k-k_\mu)|^2. \eeq

Consider a diagonal element of \eqref{eq: cfim dark field}, that is, set $\mu=\nu$, and consider the case of aberrations $\chi$ that are random on $[0,2\pi)$. In $\varphi(k,x)$, the summation over $k'$ will execute a \emph{random walk} in the Argand plane, producing an expected phase which is random on $[0,2\pi)$ (and an expected magnitude $\sqrt{\sum_{k'}|\psi_0(k-k')||\bar V(k')|}$ which has cancelled out). Hence $\varphi(k,x)$ inside the cosine in \eqref{eq: cfim dark field} is just a random phase. However, the presence of $\varphi_\mu(k,x)$ means that the phase of the term $k'=k_\mu$ is not random, which results in a \emph{biased} random walk. The degree of bias is determined by the size of $|V(k_\mu)|$ relative to the moduli of the other Fourier coefficients participating in the summation over $k'$. If $|V(k_\mu)|$ dominates the summation, as it can when the STEM objective aperture is small enough that the diffracted discs do not overlap significantly, then the ``random walk" is not random at all, and we obtain for the argument of the cosine
\beq 2\varphi_\mu(k,x) - 2\varphi(k,x) \approx 2\phi_\mu(k_\mu) - 2\phi(k_\mu) = \begin{cases} 0 & \text{for $\lambda_\mu= |V(k_\mu)|$},\\
\pi & \text{for $\lambda_\mu= \phi(k_\mu)$}. \end{cases}\eeq
Substituting into the expression for $I_{\mu\nu}^\mathrm{DF}$, we obtain
\beq\label{eq: cfim no overlap} I_{\mu\mu}^\mathrm{DF} \approx \begin{cases} 8N \sum_{|k|>K}|\psi_0(k-k_\mu)|^2 & \text{for $\lambda_\mu= |V(k_\mu)|$},\\
0 & \text{for $\lambda_\mu= \phi(k_\mu)$}. \end{cases}\eeq
In this case, we have obtained approximately full modulus information but no phase information (as expected, because this is just the classic ``phase problem" of parallel-beam diffraction). On the other hand, if $|V(k_\mu)|$ does not dominate, as is the case when the STEM objective aperture is large and multiple diffracted discs overlap significantly, then the argument of each cosine is effectively random on $[0,2\pi)$, and the cosines will tend to cancel out. In this case, we obtain
\beq I_{\mu\mu}^\mathrm{DF} \approx \begin{cases} 4N \sum_{|k|>K}|\psi_0(k-k_\mu)|^2 & \text{for $\lambda_\mu= |V(k_\mu)|$},\\
4N |V(k_\mu)|^2 \sum_{|k|>K}|\psi_0(k-k_\mu)|^2 & \text{for $\lambda_\mu= \phi(k_\mu)$}. \end{cases}\eeq
In this case, we have obtained approximately half of the modulus information and half of the phase information. We regard the latter case as the ``generic case" for 4D-STEM.

Now, still considering a diagonal element, consider the focused case $\chi=0$ (the other extreme). In this case, the phase factors involving $x$, while not random, will, when averaged over $x$, produce results very similar to those above. That is, when $|V(k_\mu)|$ dominates we obtain approximately full modulus information but no phase information, and when $|V(k_\mu)|$ does not dominate (the generic case) we obtain approximately half of the modulus information and half of the phase information.

The above findings are supported by the following table which shows numerical calculations of $I_{\mu\mu}^\mathrm{DF}$ for three different materials and three different aberration conditions (those described in the main text). The table assumes a 100~keV beam with a 20~mrad convergence semi-angle ($K=0.54$~\AA$^{-1}$). The defocused cases use $C_1=-100$~nm.  The right-hand side of the table shows the values obtained for $I_{\mu\mu}^\mathrm{DF}$ (normalized such that a value of unity means full information). Most values are close to 0.5, i.e., half of the information. Strong reflections tend to give more modulus information than phase information. The values exhibit only a weak dependence on the aberrations. These behaviors persist for higher-order reflections (not shown). COF is an acronym for covalent organic framework.

\begin{center}
\begin{tabular}{|ccccc||cc|cc|cc|}
\hline
Sample & $k_\mu$ & $d$ (\AA) & \multicolumn{2}{c||}{$V(k_\mu)$ (eV)} & \multicolumn{2}{c|}{Focused} & \multicolumn{2}{c|}{Defocused} & \multicolumn{2}{c|}{Speckled}  \\
\cline{4-5} \cline{6-7} \cline{8-9} \cline{10-11}
 &  &  & Re & Im & mod & arg & mod & arg & mod & arg\\
\hline\hline
SrTiO$_3$ [001] & $(1,0,0)$ & 3.91 & $+0.02$ & $0.0$ & 0.47 & 0.53 & 0.50 & 0.50 & 0.50 & 0.50\\
                
                & $(1,1,0)$ & 2.76 & $+6.05$ & $0.0$ & 0.55 & 0.45 & 0.54 & 0.46 & 0.55 & 0.45\\
                & $(2,0,0)$ & 1.95 & $+7.89$ & $0.0$ & 0.63 & 0.37 & 0.62 & 0.38 & 0.62 & 0.38\\
                & $(2,1,0)$ & 1.75 & $-0.15$ & $0.0$ & 0.52 & 0.48 & 0.46 & 0.54 & 0.50 & 0.50\\
                & $(2,2,0)$ & 1.38 & $+5.22$ & $0.0$ & 0.60 & 0.40 & 0.58 & 0.42 & 0.59 & 0.41\\\hline
 Graphene       & $(1,0,0)$ & 2.13 & $+1.66$ & $-2.88$ & 0.59 & 0.41 & 0.59 & 0.41 & 0.59 & 0.41\\
                & $(1,1,0)$ & 1.23 & $+2.96$ & $0.0$ & 0.71 & 0.29 & 0.70 & 0.30 & 0.71 & 0.29\\
                & $(2,0,0)$ & 1.07 & $+0.56$ & $+0.97$ & 0.63 & 0.37 & 0.62 & 0.38 & 0.61 & 0.39\\\hline
 COF-1 [001]    & $(6,\bar 3,0)$ & 2.61 & $-0.14$ & $0.0$  & 0.57 & 0.43 & 0.52 & 0.48 & 0.50 & 0.50\\
                & $(6,0,0)$ & 2.26 & $-0.69$ & $0.0$ & 0.55 & 0.45 & 0.55 & 0.45 & 0.55 & 0.45\\ 
                & $(12,\bar 6,0)$ & 1.30 & $+0.63$ & $0.0$ & 0.61 & 0.39 & 0.59 & 0.41 & 0.59 & 0.41\\
                & $(12,0,0)$ & 1.13 & $-0.22$ & $0.0$ & 0.52 & 0.48 & 0.52 & 0.48 & 0.52 & 0.48\\\hline
\end{tabular}
\end{center}

The following table includes both diagonal and off-diagonal elements of $I_{\mu\nu}^\mathrm{DF}$ for the case of a focused probe on SrTiO$_3$~[001]. $I_{\mu\nu}^\mathrm{DF}$ is a real-symmetric matrix so that values below the diagonal have been omitted. The largest off-diagonal (in terms of magnitude) is about 5 times smaller than a typical diagonal, and most off-diagonals are considerably smaller still. Note the symmetries: (i) diagonal mod-arg pairs sum to unity, (ii) off-diagonal mod-arg pairs sum to zero, and (iii) all mixed mod-arg elements are zero. These symmetries can be inferred from \eqref{eq: cfim dark field}.

\begin{center}
\begin{tabular}{|cc|cccccccccc|}
\hline
 \multicolumn{2}{|c|}{\multirow{2}{*}{SrTiO$_3$ [001]}} & \multicolumn{2}{c}{$(1,0,0)$} & \multicolumn{2}{c}{$(1,1,0)$} & \multicolumn{2}{c}{$(2,0,0)$} & \multicolumn{2}{c}{$(2,1,0)$} & \multicolumn{2}{c|}{$(2,2,0)$} \\
 & & mod & arg & mod & arg & mod & arg & mod & arg & mod & arg \\\hline
 \multirow{2}{*}{$(1,0,0)$} & mod & 0.47 & 0 & 0.003 & 0 & 0.001 & 0 & -0.06 & 0 & -0.0004 & 0  \\ 
 & arg &  & 0.53 & 0 & -0.003 & 0 & -0.0009 & 0 & 0.06 & 0 & 0.0004 \\
 \multirow{2}{*}{$(1,1,0)$} & mod &  &  & 0.55 & 0 & 0.08 & 0 & 0.001 & 0 & 0.04 & 0  \\ 
 & arg &  &  &  & 0.45 & 0 & -0.08 & 0 & -0.001 & 0 & -0.04 \\
 \multirow{2}{*}{$(2,0,0)$} & mod  &  &  &  &  & 0.63 & 0 & 0.002 & 0 & 0.05 & 0 \\
 & arg &  &  &  &  &  & 0.37 & 0 & -0.002 & 0 & -0.05\\
 \multirow{2}{*}{$(2,1,0)$} & mod  &  &  &  &  &  &  & 0.52 & 0 & 0.001 & 0 \\
  & arg &  &  &  &  &  &  &  & 0.48 & 0 & -0.001 \\
 \multirow{2}{*}{$(2,2,0)$} & mod  &  &  &  &  &  &  &  &  & 0.60 & 0\\
   & arg &  &  &  &  &  &  &  &  &  & 0.40 \\ \hline  
\end{tabular}
\end{center}

\subsection{Complete bright- and dark-field contribution}

Adding the generic dark-field component (when $|V(k_\mu)|$ does not dominate) to the bright-field component calculated earlier, we obtain the (approximate)  complete CFIM for STEM (under the WPOA)
\beq I_{\mu\mu} \approx 4N |V_\mu(k_\mu)|^2 \left( 1
- \sum_{k}|\psi_0(k-k_\mu)| |\psi_0(k+k_\mu)| \cos[2\pi(2\chi(k)-\chi(k-k_\mu)-\chi(k+k_\mu))]\right),\eeq
which is \eqref{eq: cfim stem}.

\section{Partial spatial coherence}\label{Appendix: partial spatial coherence}

\subsection{$J_{\mu\nu}$ for phase-contrast TEM}

Methods to calculate the QFIM for a mixed state are presented by Liu et al.~\cite{Liu-etal2020}. Usually, we must describe the mixed state using a density operator in Schmidt form
\beq \op\rho = \sum_\eta \eta \ketbra{\eta}{\eta}, \eeq
where $\eta$ is an eigenvalue of $\op\rho$ itself, and $\ket{\eta}$ is the corresponding eigenstate. In our case, the eigenvalues $\eta$ do not depend on the parameters, and the QFIM can be written in the form 
\beq\label{eq: qfim density matrix} J_{\mu\nu} = 4N\sum_\eta \eta\mathrm{Re}\,\braket{\eta_\mu}{\eta_\nu} - 8N \sum_{\eta,\eta'} \frac{\eta\eta'}{\eta+\eta'}\mathrm{Re}\,\braket{\eta_\mu}{\eta'}\braket{\eta'}{\eta_\nu}. \eeq 

For phase-contrast TEM, the incident density operator in Schmidt form is
\beq \op\rho_0 = \sum_\xi \tilde S(k_0) \ketbra{k_0}{k_0},\eeq
where the eigenvalue $\tilde S(k_0)$ specifies the distribution of incoherent incident plane waves (proportional to the Fourier transform of the source distribution), with normalization $\sum_{k_0}\tilde S(k_0)=1$. We make the reasonable assumption that the extent of $\tilde S(k_0)$ is much smaller than the objective aperture radius $K$. The incident density operator evolves into
\beq \op\rho = \sum_\xi \tilde S(k_0) \A(1-i\V)\ketbra{k_0}{k_0}(1+i\V)\adj\A,\eeq
which retains a Schmidt form. Using the definitions given above, we can obtain
\beq J_{\mu\nu} = 8N |V_\mu(k_\mu)|^2 \left(1-\sum_{k_0}\left(\frac{\tilde S(k_0)\tilde S(k_0+k_\mu)}{\tilde S(k_0)+\tilde S(k_0+k_\mu)} + \frac{\tilde S(k_0)\tilde S(k_0-k_\mu)}{\tilde S(k_0)+\tilde S(k_0-k_\mu)}\right)\right)\delta_{\mu\nu}, \qquad\text{$0<|k_\mu|\le K$}.\eeq
If $\tilde S(k_0)$ is a disc of radius $K_0$, then this reduces to the particularly simple form
\beq J_{\mu\nu} = 8N |V_\mu(k_\mu)|^2 \left(1- \frac{M_\mu}{M_0} \right)\delta_{\mu\nu}, \qquad\text{$0<|k_\mu|\le K$},\eeq
where $M_0$ is the number of plane waves inside the disc source, and $M_\mu$ is the number of plane waves in the overlap of two such sources displaced by $k_\mu$. Hence $M_\mu/M_0$ is just the fractional overlap of the discs. 

Thus, the effect of partial spatial coherence is to reduce the quantum Fisher information for spatial frequencies $|k_\mu|\le 2K_0$. A similar effect occurs in the case of STEM when $\chi=0$, except that here the effect occurs only at very low spatial frequencies since $K_0 \ll K$.

\subsection{$I_{\mu\nu}$ for phase-contrast TEM}

The probability of detection at a position $x$ in the image plane is $p(x)=\matel{x}{\op\rho}{x}/M$, where $\op\rho$ is the density operator given above. To calculate $p_\nu(x)$ we need
\beq\begin{split} \sum_{k_0}\tilde S(k_0) \matel{k_0}{\adj\A}{x} \matel{x}{\A\V_\nu}{k_0} &=  e^{2\pi i k_\nu\cdot x} V_\nu(k_\nu) \sum_{k_0}S(k_0) e^{2\pi i\chi(k_0) - 2\pi i\chi(k_0+k_\nu)}  + (+k_\mu\rightarrow-k_\mu) \\
&\approx  e^{2\pi i k_\nu\cdot x} V_\nu(k_\nu) e^{2\pi i\chi(0) - 2\pi i\chi(k_\nu)}\sum_{k_0} \tilde S(k_0) e^{2\pi i k_0\cdot(\nabla\chi(0) - \nabla\chi(k_\nu))}  + (+k_\mu\rightarrow-k_\mu) \\
&=  e^{2\pi i k_\nu\cdot x} V_\nu(k_\nu) e^{2\pi i\chi(0) - 2\pi i\chi(k_\nu)} S(\nabla\chi(k_\nu)-\nabla\chi(0))  + (+k_\mu\rightarrow-k_\mu),  \end{split}\eeq
where we have used a first-order Taylor expansion of the aberration function, and $S$ denotes the inverse Fourier transform of $\tilde S$. We assume a symmetric source $S$. For convenience, we define a function $s(k) \equiv  S(\nabla\chi(k)-\nabla\chi(0))$, which is real but not necessarily symmetric, and we denote its even and odd components as $s_+(k)$ and $s_-(k)$. Then we can obtain for the relevant real part
\beq \begin{split} &\sum_{k_0}\tilde S(k_0) \mathrm{Re}\{\matel{k_0}{\adj\A}{x} \matel{x}{(-i)\A\V_\nu}{k_0}\}\\
&\quad= 2 s_+(k_\nu) |V_\nu(k_\nu)| \sin[2\pi \chi(0)  -\pi \chi(k_\nu) - \pi \chi(-k_\nu) ] \cos[2\pi  k_\nu\cdot x - \pi \chi(k_\nu) +\pi \chi(-k_\nu) + \phi_\nu(k_\nu) ]\\
&\quad + 2 s_-(k_\nu) |V_\nu(k_\nu)| \cos[2\pi \chi(0)  -\pi \chi(k_\nu) - \pi \chi(-k_\nu) ] \sin[2\pi  k_\nu\cdot x - \pi \chi(k_\nu) +\pi \chi(-k_\nu) + \phi_\nu(k_\nu) ]. \end{split}\eeq
Carrying out calculations similar to the pure state case, we again find that $I_{\mu\nu}$ is diagonal. The diagonal elements can be cast into the form
\beq \begin{split} I_{\mu\mu} = 4N |V_\mu(k_\mu)|^2  \bigg(&s^2_+(k_\mu)\left( 1 - \cos[2\pi(2\chi(0) - \chi(k_\mu) - \chi(-k_\mu))]\right)\\
+ & s^2_-(k_\mu)\left( 1 + \cos[2\pi(2\chi(0) - \chi(k_\mu) - \chi(-k_\mu))]\right)\bigg),\qquad 0<|k_\mu|\le K, \end{split} \eeq
which reduces to the pure state expression on setting $s_+=1$ and $s_-=0$. Thus, the effect of partial spatial coherence is to reduce the classical Fisher information at spatial frequencies $k_\mu$ where the aberration function is varying (an anticipated result). Note that for an aberration function $\chi$ that is either symmetric or antisymmetric, we have $s_-=0$ in both cases. 

If $\tilde S(k_0)$ is a disc of radius $K_0$, 
a Zernike phase condition is obtained by choosing the symmetric aberration function
\beq \chi(k) = \begin{cases} \frac{1}{4}, & \text{$|k| < K_0$},\\
0, & \text{$|k|> K_0$}. \end{cases}
\eeq
In this case, the aberration function changes abruptly at $|k| = K_0$, so that the above assumption of a first-order Taylor expansion is invalid. However, a direct treatment of the summation $\sum_{k_0}\tilde S(k_0) e^{2\pi i\chi(k_0) - 2\pi i\chi(k_0+k_\nu)}$ is straightforward. The final result is
\beq I_{\mu\mu} = 8N |V_\mu(k_\mu)|^2 \left(1- \frac{M_\mu}{M_0} \right), \qquad\text{$0<|k_\mu|\le K$},\eeq
which is equal to $J_{\mu\mu}$. Thus, in the presence of partial spatial coherence, the Zernike phase condition enables the quantum limit for spatial frequencies admitted by the objective aperture.

\subsection{$J_{\mu\nu}$ for 4D-STEM}

Each of the $M$ independent quantum systems is now in a mixed state, and the appropriate tensor product state is
\beq \op\rho = \op\rho(x_1)\otimes\cdots\otimes\op\rho(x_M). \eeq
Calculation of $J_{\mu\nu}$ via expression \eqref{eq: qfim density matrix} requires each $\op\rho(x)$ in diagonal form, which is a challenging problem. We will rather examine the spatially \emph{incoherent} case, and infer the partially coherent case via interpolation. The incoherent case was, in fact, calculated above for phase-contrast TEM. Here the result becomes
\beq J_{\mu\nu} = 8N |V_\mu(k_\mu)|^2 \left(1- \frac{M_\mu}{M_K} \right)\delta_{\mu\nu},\eeq
where $M_\mu/M_K$ is the fractional overlap of the bright-field disc and the disc centered at $k_\mu$. The spatial frequency $k_\mu$ is unrestricted. This result is similar to the pure state expression except that here there is no possibility of tuning the autocorrelation owing to the incoherence. 

We infer by interpolation that, even for optimum tuning of the aberrations, partial spatial coherence will permit only an incomplete reduction of the autocorrelation term. Thus, there is some reduction of the 4D-STEM quantum Fisher information for spatial frequencies $|k_\mu|\le 2K$. Quantum Fisher information for spatial frequencies $|k_\mu| > 2K$ is unaffected.

\subsection{$I_{\mu\nu}$ for 4D-STEM}

Again, we will infer the result by interpolating between the pure state case and the incoherent case. Using manipulations similar to those already provided in detail, we find that the CFIM for the incoherent case contains only modulus information (as expected):
\beq I_{\mu\nu} = 8N|V(k_\mu)||V(k_\nu)|\sum_{|k|>K}\frac{|\psi_0(k-k_\mu)|^2|\psi_0(k-k_\nu)|^2}{\sum_{k'}|\psi_0(k-k')|^2|V(k')|^2}, \qquad \lambda_\mu=|V(k_\mu)|, \lambda_\nu=|V(k_\nu)|. \eeq
Moreover, this CFIM is non-diagonal, and it consists solely of dark field contributions (the bright field contributions vanish). If there is no overlap of the diffraction discs, then it reduces to \eqref{eq: cfim no overlap}, i.e., full modulus information, as it should.

We infer by interpolation that partial spatial coherence reduces those elements of the 4D-STEM CFIM that refer to the phases, which occurs for all spatial frequencies $k_\mu$. For CFIM elements that refer to the moduli, if they are comprised mostly of bright field contributions then they are reduced, whereas if they are comprised mostly of dark field contributions then they are possibly increased.

\end{widetext}

%

\end{document}